\documentclass[letter,twocolumn]{jpsj3} 
\usepackage{times}

\def\lsim{\lower -0.3ex \hbox{$<$} \kern -0.75em \lower 0.7ex \hbox{$\sim$}}

\def\jo #1#2#3#4{#1 {\bf #2} (#3) #4}

\def\PRB{Phys.\ Rev.\ B}

\def\JPSJ{J.\ Phys.\ Soc.\ Jpn.}

\def\CM{Chem.\ Mater.}
\def\EPJB{Eur.\ Phys.\ J.\ B}
\def\STAM{Sci.\ Technol.\ Adv.\ Mater.}

\usepackage{bm} 

\def\runauthor{H. Yoshioka \textit{et al.}}

\title{Incommensurate Antiferromagnetic Insulating State in 
(MDT-TS)(AuI$_2$)$_{x}$}

\author{Hideo \textsc{Yoshioka}$^{1}$\thanks{E-mail: h-yoshi@cc.nara-wu.ac.jp}, Hitoshi \textsc{Seo}$^{2,3}$\thanks{E-mail: seo@riken.jp}, and Yuichi \textsc{Otsuka}$^{2,3}$\thanks{E-mail: otsukay@riken.jp}}

\inst{$^{1}$Department of Physics, Nara Women's University, Nara
630-8506 \\
$^{2}$Condensed Matter Theory Laboratory, RIKEN, 2-1 Hirosawa, Wako,
Saitama 351-0198 \\
$^{3}$JST, CREST, 2-1 Hirosawa, Wako, Saitama 351-0198}

\recdate{\today}

\abst{
We theoretically study 
 the metal-insulator transition in a molecular conductor (MDT-TS)(AuI$_2$)$_{x}$ 
composed with
an incommensurate 
ratio ($x = 0.441$), 
 where the conduction band 
originated from 
 the HOMO of donor MDT-TS molecules 
is incommensurately filled.  
We consider a 
 two-dimensional 
Hubbard model taking account of 
anisotropic transfer integrals in the donor layer, 
 under a periodic potential due to the anions (AuI$_2$)$^-$ 
 which mismatches the donor lattice period, 
 and investigate the ground state within 
mean-field approximation.
An antiferromagnetic insulating state with induced charge disproportionation is
 obtained
 in the large $U$ region; 
this corresponds 
to the incommensurate Mott insulating state 
 predicted previously 
 [H. Yoshioka {\it et al.}: \jo{\JPSJ}{74}{2005}{1922}]
 based on a simplified one-dimensional model. 
}

\kword{molecular conductors, Hubbard model, incommensurate filling, metal-insulator
transition, Mott insulator, organic conductors}

\begin{document}
\maketitle

In recent years, molecular conductors with incommensurate (IC) composition ratios have been synthesized 
 and their electronic properties are explored experimentally.~\cite{TMori_ChemRev,Kawamoto_2009STAM}
Among them, a series of isostructural compounds expressed as $AB_x$ 
 where $A$ 
denote donor molecules such as 
 MDT-TSF, MDT-ST, MDT-TS, 
 and $B$ are 
 monovalent anions taking (AuI$_2$)$^-$ and (I$_3$)$^-$,
 have been systematically investigated~\cite{Takimiya_2001ACIE,Kawamoto_2002PRB,Takimiya_2003CM_1,Takimiya_2003CM_2,Kawamoto_2003PRB,Kawamoto_2003EPJB,Takimiya_2004CM,Kawamoto_2005PRB_1,Kawamoto_2005PRB_2,Kawamoto_2005JPSJ,Kawamoto_2006PRB_1,Kawamoto_2006PRB_2,Kawamoto_2008PRB}. 
They form quasi-two-dimensional crystal structures with 
 alternatively stacking $A$ and $B$ layers.  
Since $B$ takes closed shell configurations, 
 the irrational values of $x \simeq 0.41-0.45$ result in irrational average valences 
 for the donors $A^{+x}$;
 then the conduction band 
 composed of the HOMO of donors 
becomes incommensurately filled.
The filling factors are $n_e=2-x$, 
 slightly larger than $3/4$-filling ($n_e=3/2$) 
 which would correspond to $x=1/2$ (i.e., $A_2B$); 
 in the hole picture, they are slightly smaller than quarter-filling.   

As is naturally expected from such IC band fillings, 
 most of these compounds show metallic behavior,
 and some undergo a superconducting transition with a maximum $T_c \simeq 4$~K.\cite{Takimiya_2001ACIE,Kawamoto_2002PRB,Takimiya_2003CM_1,Takimiya_2003CM_2,Kawamoto_2005JPSJ,Kawamoto_2006PRB_2}
However, exceptionally, (MDT-TS)(AuI$_2$)$_x$ ($x =0.441$) 
shows a metal-insulator crossover at $T_\rho = 85$~K 
where 
the temperature-dependence of the resistivity has a minimum.~\cite{Takimiya_2004CM,Kawamoto_2005PRB_1,Kawamoto_2008PRB} 
With further decreasing the temperature,
an antiferromagnetic (AF) transition takes place at $T_{\rm N} = 50$~K. 
By applying external pressure, $T_\rho$ decreases
and a superconducting phase appears 
above
$P_c = 10.5$~ kbar. 
The fact that the insulating state does not 
 accompany a magnetic order in the temperature range $T_\rho > T > T_{\rm N}$
 implies that the insulating behavior in this compound 
 might be due to strong correlation, 
 in spite of the IC filling where strong-coupling insulators are usually not expected.  
In contrast, 
 Mott insulators and charge ordered insulators are frequently observed
 in typical molecular conductors with commensurate fillings, 
 which are theoretically studied based on Hubbard-type models extensively~\cite{Seo_JPSJReview}. 

A key to understand the origin of the insulating behavior in (MDT-TS)(AuI$_2$)$_x$  
 is the fact that the anions $B^-$ in these materials are 
 not randomly distributed: 
X-ray scattering experiments show that they form regular lattices
with a different periodicity from that of donor
 lattices.\cite{Kawamoto_2002PRB,Takimiya_2003CM_1,Takimiya_2003CM_2,Kawamoto_2005PRB_2}               
Owing to this
mismatch in periodicities,
the anions are expected 
to influence the electronic state in the donor HOMO band
as a periodic potential. 
In fact, reconstructions of the Fermi surface by the anion potential 
in metallic compounds are observed 
by the Shuvnikov-de Haas oscillations.~\cite{Kawamoto_2003PRB,Kawamoto_2006PRB_1}  

\begin{figure}[b]
\begin{center}
 \includegraphics[width=6cm,clip]{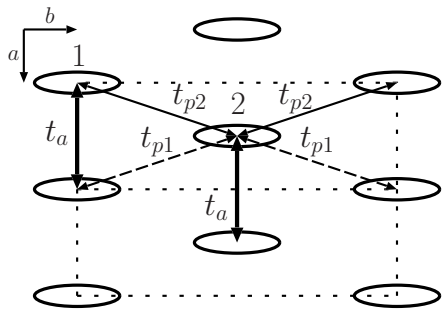}
\end{center} 
\caption{
Schematic representation of the two-dimensional donor 
layer in (MDT-TS)(AuI$_2$)$_x$. 
Ellipses denote MDT-TS molecules 
 and the unit cell written by the dotted lines contains two equivalent molecules. 
The 
transfer integrals 
$t_a$, $t_{p1}$, and $t_{p2}$ are 
represented by 
arrows. 
}
\label{fig:2d_MDT_model}
\end{figure}  

Taken these facts together,
 a peculiar Mott insulating state in 
electronic 
 systems with IC band
 filling has been theoretically proposed~\cite{Yoshioka_2005JPSJ,SeoProc}. 
When the periodic anion potential is strong enough, 
 the conduction band becomes effectively half-filled, 
 since the filling does match 
 the periodicity of monovalent anions that provide
 one hole per anion to the donors. 
As a result, the mutual Coulomb interaction between electrons 
 on the donor lattice 
is expected to lead  
to a Mott insulating state. 
Its existence was 
confirmed within 
a simplified one-dimensional (1D) model 
which incorporates the IC filling 
as well as the anion potential, 
by using bosonization and perturbative renormalization group methods~\cite{Yoshioka_2005JPSJ}.
This strong-coupling insulating state, 
 which is produced by the cooperative effect 
 of anion potential and electronic correlation, 
 is called as ``IC Mott insulator''
in order to distinguish it from the usual Mott insulator
at half filling. 
In the former, the localized spins do not situate on the donor sites in a commensurate way 
 but follow the periodicity of the anion potential; 
 in the latter spin $1/2$ simply localizes on each site. 

In the present paper, 
 beyond the simplified 1D model,
 we study 
 a two-dimensional Hubbard model with 
 anisotropic transfer integrals between the donors 
 which reflect the actual structure of (MDT-TS)(AuI$_2$)$_x$.
Within
 mean-field approximation,
 we obtain 
 self-consistent insulating solutions at large Coulomb interaction, 
considered as  
the IC Mott insulator, 
 where the charge disproportionation induced by the anion potential
 coexists with AF ordering.   

The model for (MDT-TS)(AuI$_2$)$_x$ we consider is represented as 
\begin{align}
 H &=  \sum_{\langle ij \rangle,\sigma} t_{ij} \left(c_{i,\sigma}^\dagger c_{j,\sigma} +
 \mathrm{h.c.} \right) \nonumber \\ 
 & \hspace{1cm} + U \sum_{i} n_{i,\uparrow} n_{i,\downarrow} +
 \sum_{i} v_i n_i, 
\end{align}
where $c_{i,\sigma}^\dagger$ is the creation operator of the electron with
spin 
$\sigma = \uparrow, \downarrow$ 
at the $i$-th donor site. 
The crystal structure 
and
the hopping parameters $t_{ij}$
in the donor layer 
are 
schematically shown in Fig.\ref{fig:2d_MDT_model}. 
We set values of $t_{ij}$ as those calculated
by the extended H$\ddot{\rm u}$ckel method,~\cite{Kawamoto_2005PRB_1}  
$t_a = 76.0$ meV, $t_{p1} = -7.9$ meV, and $t_{p2} = -26.3$ meV, 
and consider the on-site Coulomb interaction $U$ for each donor site. 

We assume that the potential from the anions, expressed as $v_i$,  
depends only on the coordinate
 along the $a$-axis 
 in the respective columns,  
 considering the fact that in the actual 
materials  
 the anions
 align in this direction~\cite{Takimiya_2003CM_1,Takimiya_2003CM_2,Takimiya_2004CM,Kawamoto_2005PRB_2}. 
Namely, we 
adopt the form: 
\begin{align} 
 v^{(1)}_i &= \delta \cos\left(2 \pi x \xi^{(1)}_i \right) \\ 
 v^{(2)}_i &= \delta \cos\left(2 \pi x \left\{\xi^{(2)}_i-1/2+\phi\right\}\right), 
\end{align}
where $v^{(\mu)}_i$ and $\xi^{(\mu)}_i$ ($\mu=1,2$) express 
 the anion potential and the coordinate along the $a$-axis for 
 molecule $i$ 
 in the $\mu$-th column 
 of the unit cell; 
 we take the intermolecular distance in this direction as $a=1$. 
The strength of the anion potential is denoted by $\delta$.
Note that, from the chemical formula (MDT-TS)(AuI$_2$)$_x$, 
 the anion periodicity is given by $1/x$. 
The relative position between the two anion columns, 
 which determine the phase $\phi$,
 is taken as 0 in this work,
 but the overall features below 
 remain unchanged for different values of $\phi$. 
Since the anions become monovalent 
 by subtracting electrons from the donor layer,
 the filling factor of the donor sites in terms of holes, $n_h=2-n_e$, 
 coincides with  the inverse of the anion periodicity, 
 namely, $x$. 
In the following, we take the hole picture. 

The on-site Coulomb interaction term 
 is treated within the mean-field approximation as 
$n_{i\uparrow}n_{i\downarrow} \simeq 
  \langle n_{i\uparrow} \rangle n_{i\downarrow} 
+ n_{i\uparrow} \langle n_{i\downarrow} \rangle
- \langle n_{i\uparrow} \rangle \langle n_{i\downarrow} \rangle$ 
and self-consistent solutions are 
obtained
at $T=0$ 
for 
$\langle n_{i \sigma} \rangle$ $(\sigma=\uparrow, \downarrow)$ 
on all sites in the supercell we take; 
 we do not assume any functional form, 
 e.g., a sinusoidal function for the charge and spin densities.  
We consider several choices for the anion periodicity $1/x$, 
 such as, $x=1/3$, $2/5$, $3/7$, and $4/9$, 
  which are close to $x=0.441$ but, to be exact, rational.
This is due to a practical limitation that
 the size of the supercell has to be finite.
We choose the supercells as 
 6, 5, 14, and 9 times  the unit cell along the $a$-axis 
 for the choices 
 of $x$
 above, respectively, 
 while that along the $b$-axis is unchanged. 
The main conclusions we reach below 
 do not depend on the choice of $x$,
 therefore we expect that 
 our results
 hold for purely IC systems as well.
\begin{figure*}[htb]
\begin{center}
 \includegraphics[width=5cm,clip]{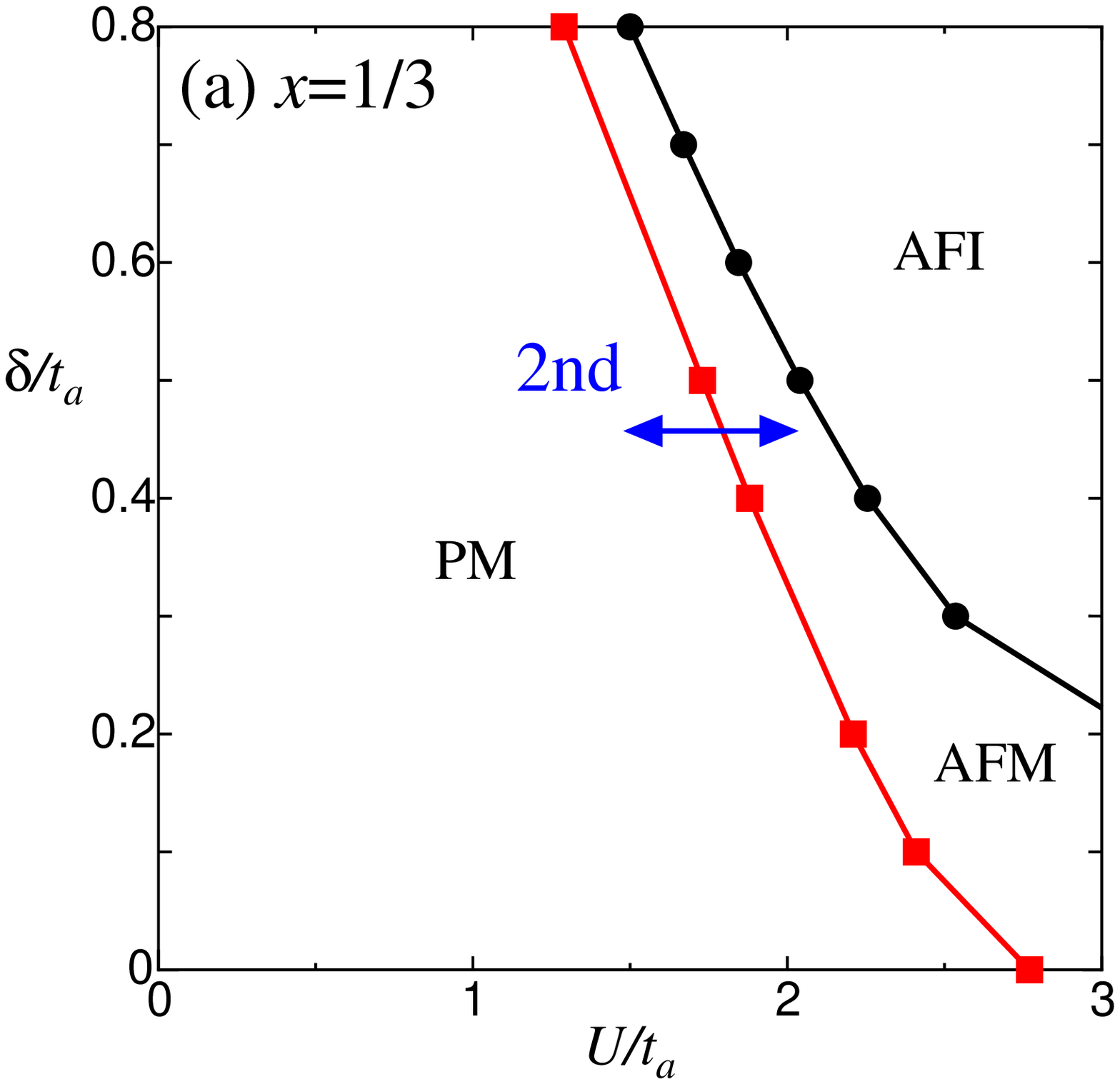}
 \includegraphics[width=5cm,clip]{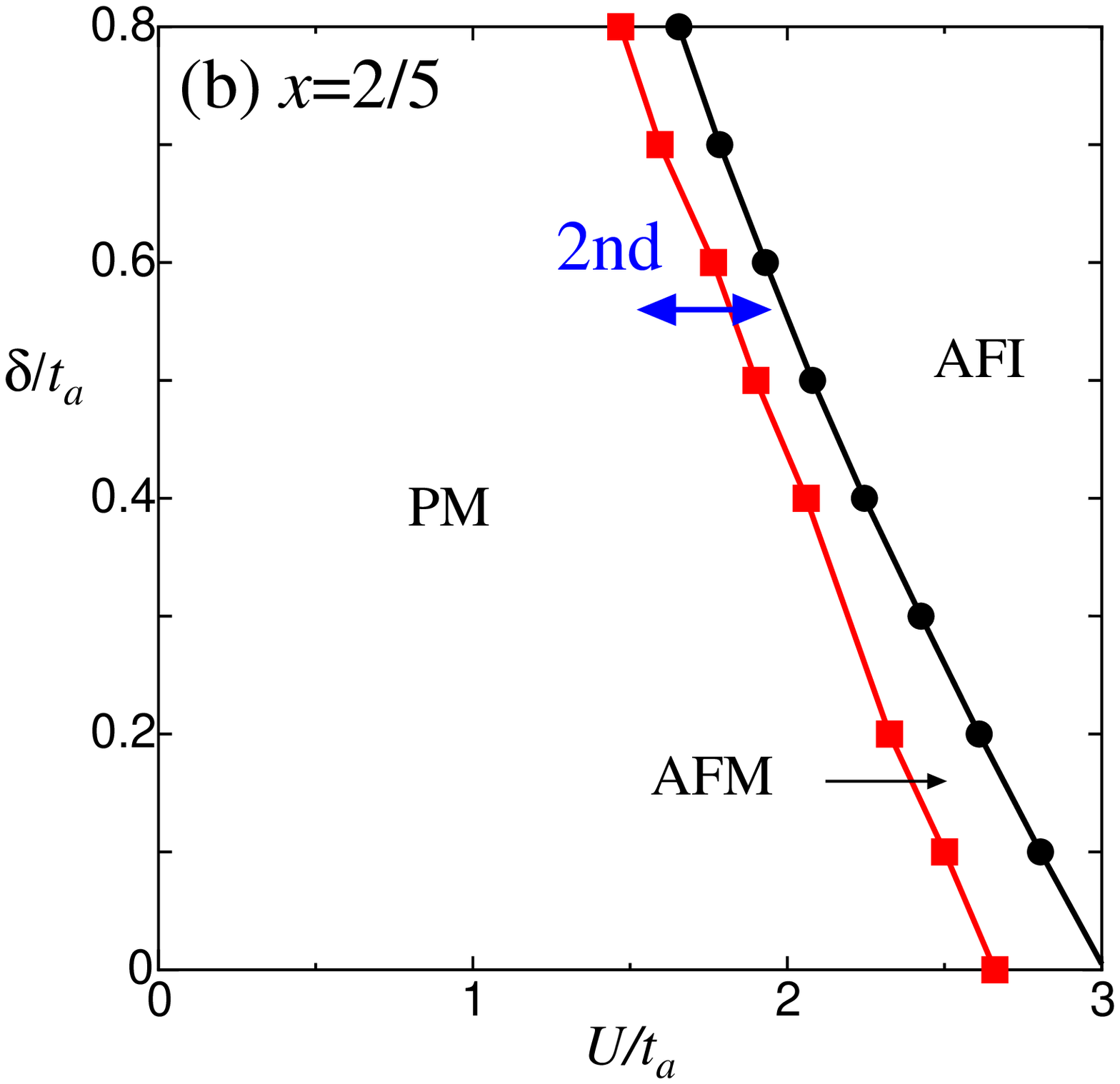} \\
 \includegraphics[width=5cm,clip]{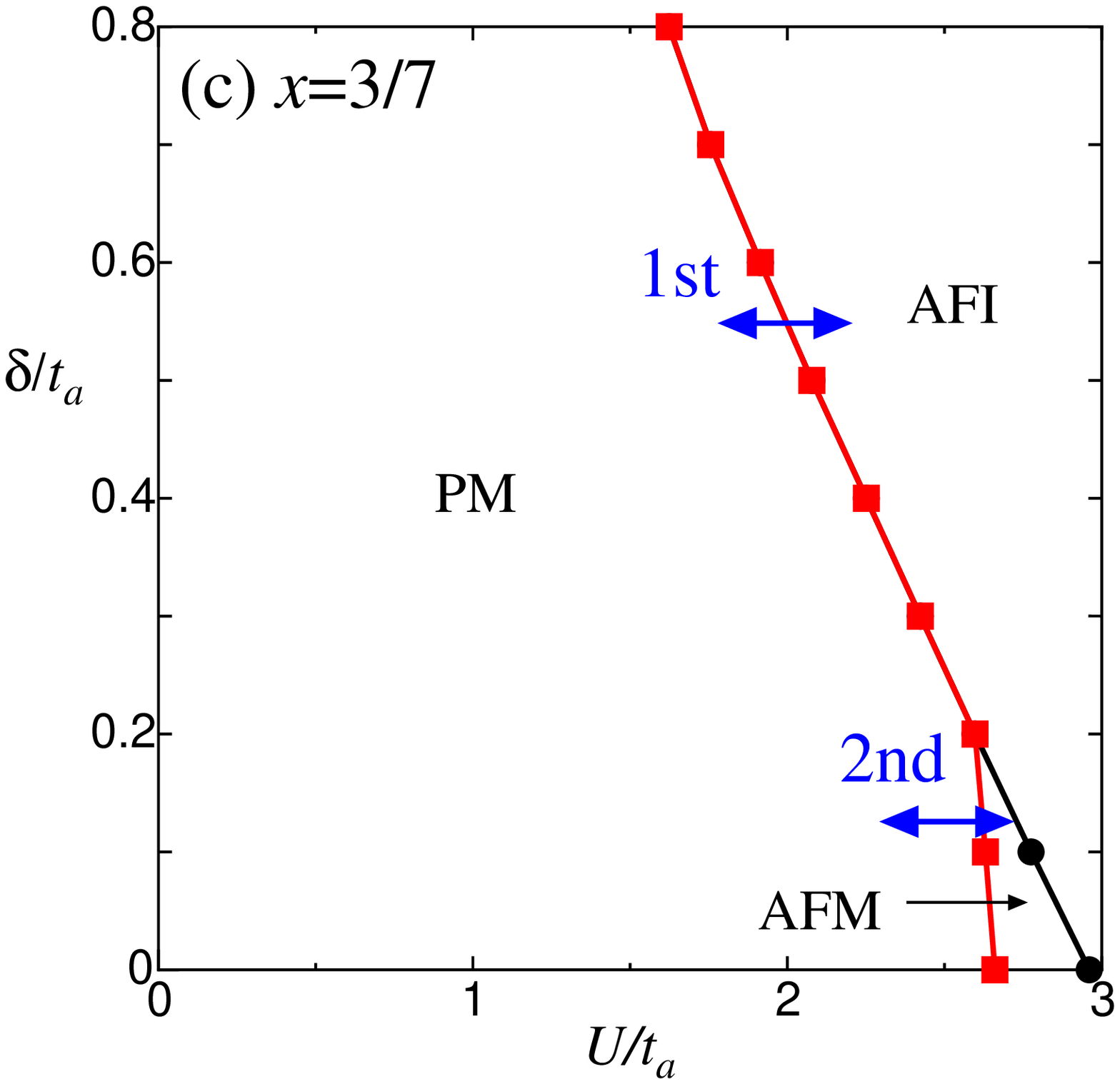}
 \includegraphics[width=5cm,clip]{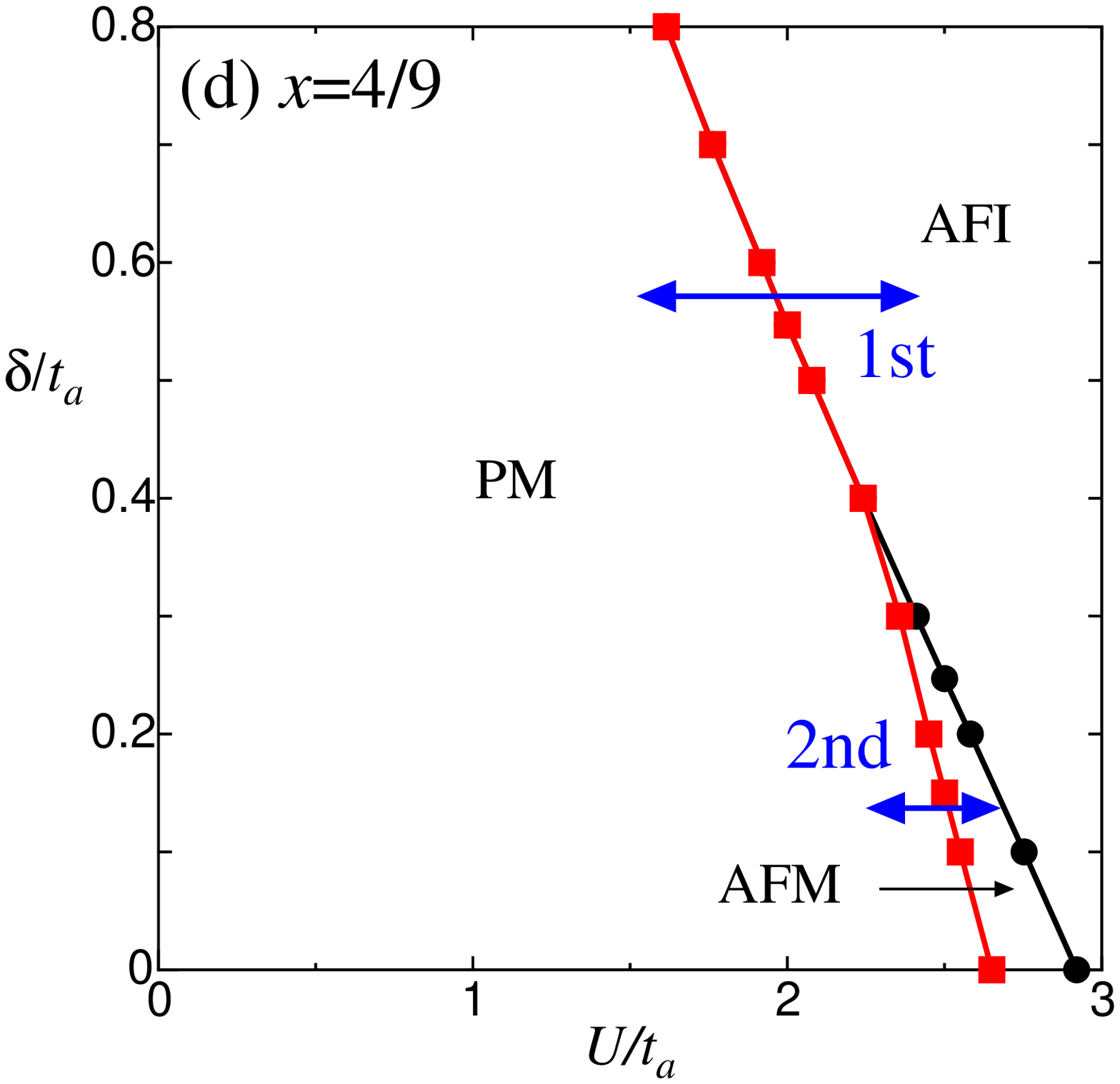}
\end{center} 
\caption{(Color online) 
Ground-state 
phase diagrams obtained by the mean-field approximation for 
 the two-dimensional model for 
 (MDT-TS)(AuI$_2$)$_x$ with $x=1/3$ (a), $x=2/5$ (b), $x=3/7$ (c), and $x=4/9$ (d). 
$\delta$ and $U$ are the strength of the anion potential and the
 on-site Coulomb repulsion. 
PM, AFM, and AFI denote the paramagnetic metallic, antiferromagnetic metallic, and 
 antiferromagnetic insulating phases, respectively. 
The paramagnetic-antiferromagnetic and metal-insulator phase boundaries are 
 shown by a red curve with squares and by a black curve with circles, respectively,
 while ``1st'' (``2nd'') expresses the first (second) order phase transition.    
}
\label{fig:2d_MDT_phase}
\end{figure*}

In Fig. \ref{fig:2d_MDT_phase}, 
 the ground-state phase diagrams for 
the 
different values of $x$ 
are shown, 
 on the plane of 
the strength of the anion potential $\delta$ and the on-site Coulomb repulsion $U$, 
 both normalized by the largest transfer integral $t_a$, 
 that along the $a$-axis.  
In all cases, three distinct phases are obtained.  
The paramagnetic metallic (PM) phase exists for relatively weak $\delta$
and $U$, 
whereas the  antiferromagnetic insulating (AFI) state is realized
for strong $\delta$ and $U$.
Between 
these two states, 
 there appears an AF metallic (AFM) phase  
 in which the Fermi surface partially remains even in the presence of the AF order.
The transition between the PM and AFM phases is continuous, 
 while a first order phase transition occurs between the PM and AFI phases, 
 for all cases of $x$. 
 
All the phase diagrams 
 share common feature that
 the reduction of the band width 
 and/or the enhancement of the anion potential 
 leads to the insulating state: 
 the on-site Coulomb repulsion $U$ in the donor layer 
 and the anion potential $\delta$ act cooperatively 
 to give rise to the AFI state.
This is 
 analogous to the results in 
 the simplified 1D model~\cite{Yoshioka_2005JPSJ}, 
 where the 
 IC Mott insulator is stabilized when both $U$ and $\delta$ are large; 
 however, 
 the strong fluctuation, 
 which is significant in 1D, 
 prevents the AF long-range order,
 and then its phase diagram has only two phases. 
It should be noted that 
 effects of quantum fluctuation are neglected
 in the present mean-field treatment, 
 so the regions of the AF ordering are overestimated. 

\begin{figure}
\begin{center}
\hspace{-1em}
 \includegraphics[width=7.5cm,clip]{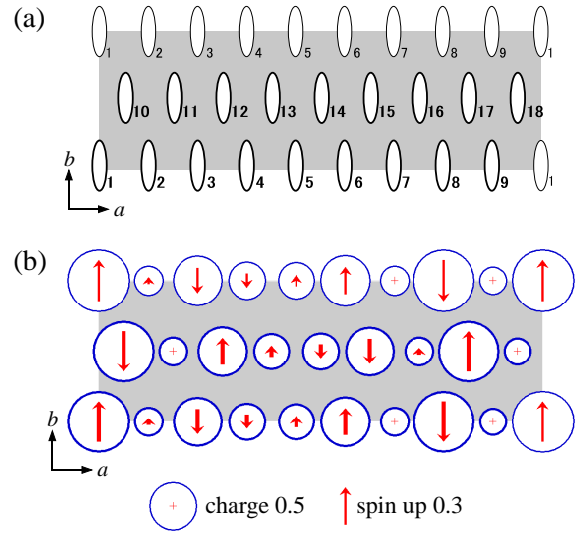}
\end{center} 
\caption{(Color online) 
Supercell (a) and 
 the distribution of charge and spin densities on each site
 in the mean-field solution 
(b) 
 for $x=4/9$. 
The gray rectangle is the supercell 
including independent 18 donor sites.
Each sites is distinguished by the index shown in (a), 
which notation is also used in Figs.~\ref{fig:n49-U2.0} and~\ref{fig:n49-U2.5}.
In (b), 
the result for $U/t_a = 1.9$ and $\delta/t_a = 0.7$ is shown; 
 the amount of charge density is expressed by the size of the circle,
 whereas the direction and the length of the arrow 
 indicate
 up/down and the magnitude of spin moment.
 Reference scales are shown in the bottom.
}
\label{fig:n49-unitcell}
\end{figure}  

\begin{figure}
\begin{center}
 \includegraphics[width=6.5cm,clip]{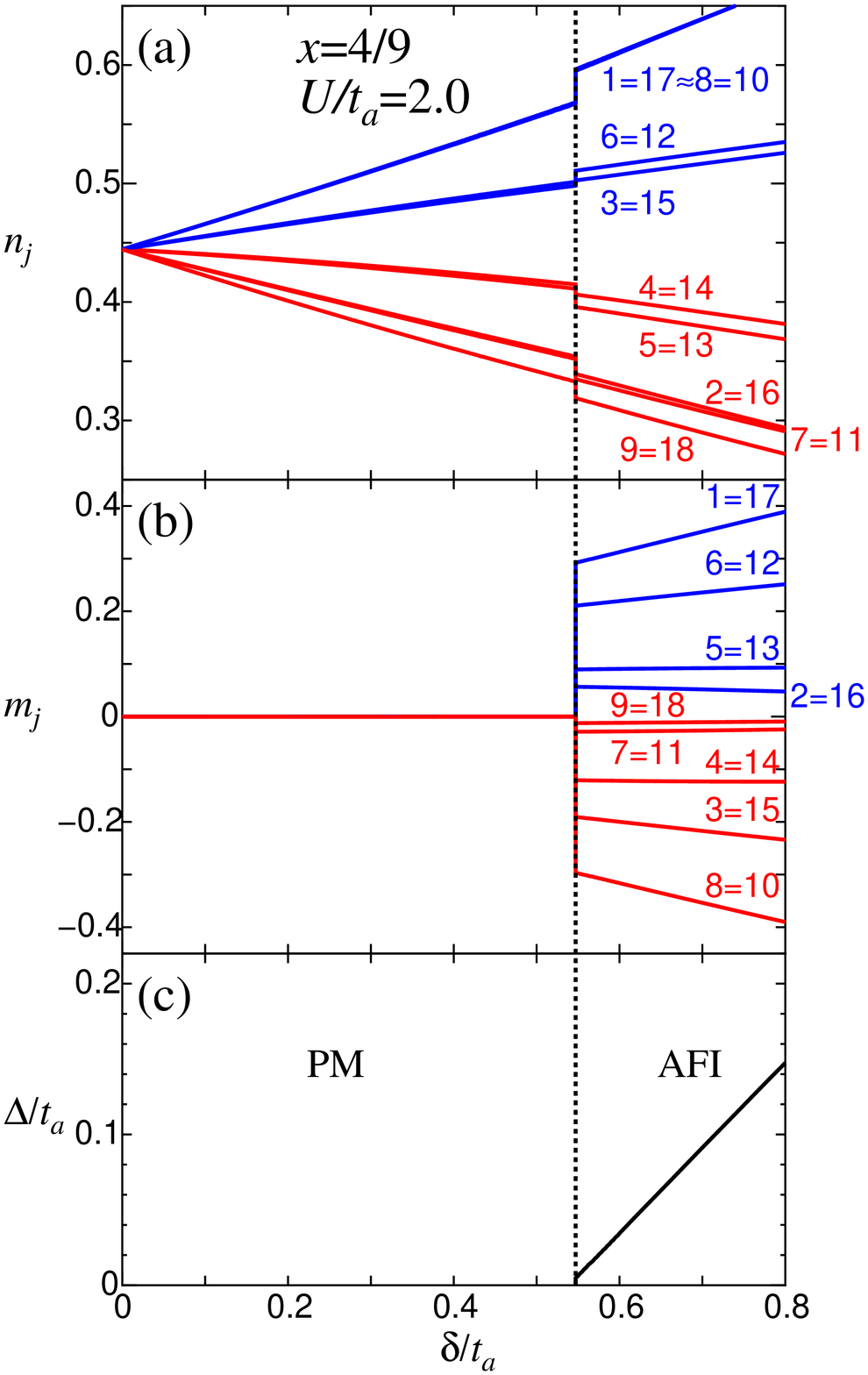}
\end{center} 
\caption{(Color online) 
Charge (a) and spin (b) distribution and the charge gap $\Delta$ (c) 
for 
the case of $x=4/9$, 
 as a function of $\delta/t_a$  for fixed $U/t_a = 2.0$. 
The integers indicate site indices in the supercell as shown in
 Fig. \ref{fig:n49-unitcell}.   
}
\label{fig:n49-U2.0}
\end{figure} 

\begin{figure}
\begin{center}
 \includegraphics[width=6.5cm,clip]{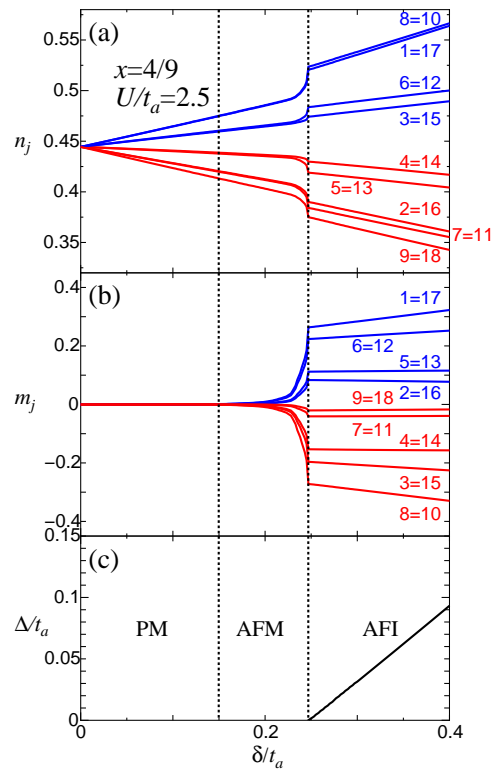}
\end{center} 
\caption{(Color online) 
Charge (a) and spin (b) distribution and the charge gap $\Delta$ (c) 
for 
the case of $x=4/9$, 
 as a function of $\delta/t_a$  for fixed $U/t_a = 2.5$. 
The integers indicate site indices in the supercell as shown in
 Fig. \ref{fig:n49-unitcell}.   
}
\label{fig:n49-U2.5}
\end{figure}

We next 
show the distribution of charge and spin densities,
 in the case of $x=4/9$,
 where there exist 18 donors and 8 carrier holes 
 in the supercell [see Fig. \ref{fig:n49-unitcell} (a)]: 
The $\delta$-dependences of 
 the charge and spin densities at each site together with the charge gap,  
 for $U/t_a = 2.0$ and $U/t_a = 2.5$ are shown in
 Figs. \ref{fig:n49-U2.0} and \ref{fig:n49-U2.5}, respectively.
 
In the 
former case, 
 the PM and AFI states are divided
 by the sharp first order transition.
In the PM state, as seen in Fig. \ref{fig:n49-U2.0}(a),
the charge on each site is always 
disproportionated according to the anion potential~\cite{charge}.
When the first order transition sets in,
the charge disproportionation is further enhanced;
however, 
the symmetry in the charge sector 
is not lowered.
The spin arrangement in the AF state 
 can be understood as the Neel ordering
 of the spins localized at the charge rich sites: 
 in Fig. \ref{fig:n49-unitcell} (b), the real-space 
 configuration of the charge and spin pattern is shown.
One can regard this pattern as 
 effective $S= 1/2$ localized spins
 on each $9/4$ molecules, 
which is the anion periodicity $1/x$, 
 forming a simple staggered AF pattern 
among 8 spins in the supercell. 

In the case of $U/t_a = 2.5$ (Fig.~\ref{fig:n49-U2.5}), 
on the other hand, 
 as increasing the anion potential 
 the ground state first changes from PM to AFM,
 and then to the AFI state. 
As in the case 
of 
$U/t_a = 2.0$, the charge disproportionation 
exists even in the PM state and it is enhanced as $\delta$ is increased. 
The charge distribution does not show any
 clear singularities at the transition between PM and AFM. 
The charge pattern itself is always 
 the same in the whole phase diagram; 
 this holds for 
 each of 
the $x$ values we considered, 
 namely, for all the phase diagrams shown in Fig.~\ref{fig:2d_MDT_phase}.

The 
AFI state we obtain can be interpreted as 
 the IC Mott insulator 
 which was predicted 
 in the 1D model~\cite{Yoshioka_2005JPSJ}, 
with which the magnetic ordering is accompanied in the present case.  
The periods of the charge and the spin densities do not match 
 the donor lattice where the carriers exist, 
 but follow the anion periodicity without changing the 
 charge pattern from the PM state. 
These features are expected from 
 the previous results 
 in the 1D case: 
 the charge is arranged according to the anion potential, 
 and then the period of charge arrangement is 
 IC to that of donors where the carriers exist. 
The translational symmetry is not broken in the transition to  
 the IC Mott insulating state.
The spin degree of freedom behaves as a Heisenberg localized spin, 
 therefore as a spin liquid in the 1D model, 
 which is expected to show AF order when the dimensionality 
 effect is added, as in fact realized in the present calculation. 

Let us note that the IC Mott insulating state should be
distinguished from the charge ordered state 
widely observed in the 
$A_2B$-type 
molecular conductors. 
In the latter, the charge disproportionation occurs due to 
the long range component of the mutual interaction and then
the translational symmetry is broken. 
We also note another difference between the 1D case and our result for
the 2D case: 
The AFI state is stabilized in the 2D model down to $\delta=0$, 
 whereas in the 1D model finite $\delta$ is needed to 
give rise to the IC Mott insulator. 
The present result in the 2D case suggests the existence of nesting property 
 stabilizing AF, but we cannot ignore the possibility that 
 it is an artifact of our mean-field treatment; 
 as mentioned above, mean-field approximation usually overestimates ordered states. 
Calculations incorporating quantum effects are left for future studies; 
our result that 
 the IC AFI state is stabilized 
when $\delta$ and $U$ are large  
presumably 
remains unchanged. 

The obtained phase diagrams are consistent with the experimental results that 
applying pressure leads to the transition from the Mott insulating state 
to the metallic state~\cite{Kawamoto_2005PRB_1}, 
since the pressure is expected to increase the bandwidth. 
An interesting problem to be explored would be whether the IC Mott insulating state 
 survives just up to the superconducting phase, 
 similarly to the half-filled cases such as in $\kappa$-ET$_2X$ system. 
Another possibility is that a spin-density-wave phase appears 
 between the IC Mott insulating and the superconducting phases, 
 typically 
observed in quasi-1D systems such as TM$_2X$.
The nesting induced spin-density-wave state 
might correspond
to 
AF states that we obtained in the small $\delta$ and/or $U$ region, 
 although at the mean-field level we cannot distinguish it 
from the AF ordering due to the Mott insulating state.

In summary, we theoretically investigated the ground-state properties 
of (MDT-TS)(AuI$_2$)$_{0.441}$ 
by studying a two-dimensional Hubbard model under a periodic potential
due to the anions 
within mean-field approximation. 
Three kinds of states, 
PM, AFM, and AFI states were obtained, 
and the AFI state is stabilized when both $U$ and $\delta$ are strong. 
This state 
is analogous to 
the incommensurate Mott insulating state 
 obtained 
previously
by the bosonization and renormalization group study in a one-dimensional model~\cite{Yoshioka_2005JPSJ}. 

\section*{Acknowledgment}
The authors would like to thank H. Fukuyama for valuable comments. 
This work was supported by Grant-in-Aid for Scientific Research
(Nos. 20110002, 20110004, 21740270 and 22540329) from the Ministry of Education, Culture, Sports, Science
and Technology 
and by Nara Women's University Intramural Grant for Project Research.


\begin{thebibliography}{99}
\bibitem{charge}
Although the charge distribution is mostly determined by the anion potential on each site,  
 the anisotropy in the transfer integrals as well as the anion potential on the surrounding sites 
 also affect it, therefore even for the sites where the on-site potential is equal 
 deviations of the charge density occur (for example sites ``3" and ``6" in Fig.~\ref{fig:n49-U2.0}(a)).
\end{thebibliography}
\end{document}